\documentclass[%
 aip,
 amsmath,amssymb,
reprint,%
]{revtex4-1}
\usepackage{float} 
\usepackage{graphicx}
\usepackage{dcolumn}
\usepackage{bm}

\usepackage[utf8]{inputenc}
\usepackage[T1]{fontenc}
\usepackage{mathptmx}
\usepackage{etoolbox}

\makeatletter
\def\@email#1#2{%
 \endgroup
 \patchcmd{\titleblock@produce}
  {\frontmatter@RRAPformat}
  {\frontmatter@RRAPformat{\produce@RRAP{*#1\href{mailto:#2}{#2}}}\frontmatter@RRAPformat}
  {}{}
}%
\makeatother
\begin{document}

\preprint{AIP/123-QED}


\title{Ionization rate vs.\ laser intensity determined from ion count vs.\
peak intensity due to neutral gas exposure to an 800~nm ultrashort pulsed
laser}
\author{E. L. Ruden}
\affiliation{Air Force Research Laboratory, Directed Energy Directorate}
\keywords{Ultrashort pulsed laser, above threshold ionization rate}

\begin{abstract}
The optical cycle-averaged ionization rate of Ar, O$_{2}$, and N$_{2}$ vs.\
local instantaneous laser intensity $I$ for linear polarized $800$~nm light
is determined up to approx.\ $300$ TW/cm$^{2}$ by numerically inverting
published time-of-flight ion spectrometer data. The published Ar$^{+}$
collection efficiency of the microchannel plate (MCP) at the end of the
spectrometer and its $I_{0}$ scale are recalibrated by fitting it to its
high $I_{0}$ solution. The relative collection efficiencies of the other
species are determined by published MCP cathode data. Results for O$_{2}$
are consistent with a reevaluation of published data used to determine its
cross section $\sigma _{8}$\ in the multiphoton (low $I$) regime.
\end{abstract}
\volumeyear{year}
\volumenumber{number}
\issuenumber{number}
\eid{identifier}
\date[Date text]{date}
\received[Received text]{date}

\revised[Revised text]{date}

\accepted[Accepted text]{date}

\published[Published text]{date}

\startpage{1}
\endpage{2}
\maketitle

\section{Introduction}

This is one of two complementary papers. It and the other \cite{Ruden25QmodT}
are used by a third \cite{Ruden25QmodS} to determine the electron density,
electron temperature, and internal laser intensity of filaments formed in
air by a linearly polarized Ti:sapphire ultrashort pulsed laser (USPL) with
wavelength $\lambda =800$~nm, based on measurements of the filaments'
electrical conductivity (a function of both density and temperature). The
results, though, are of intrinsic interest for many USPL applications.

We determine here the cycle-averaged ionization rate $\left\langle
W\right\rangle _{\gamma }$ vs.\ local instantaneous laser intensity $I$ by
inversion of the ion counts per shot-Torr $\kappa N\left( I_{0}\right) $
measurements of Guo, et al.\ \cite{Guo98} (\textquotedblleft
Guo\textquotedblright ) recorded by a microchannel plate (MCP) at the end of
a time-of-flight (TOF) spectrometer. For brevity, published references will
generally be referred to by the last name of the first author after it has
been fully cited. $N\left( I_{0}\right) $ here is the number of the majority
ion species produced (the neutral molecule with a single electron removed)
by a laser pulse of peak laser intensity $I_{0}$ after its passage through a
very low density neutral gas within an aperture-limited collection volume.
We take this to represent the electron number, neglecting other ion species
contributions. This is generally a good approximation, except for the
highest values of $I_{0}$ considered, for which our inversion method fails
regardless due to excessive depletion of the neutral species. $\kappa $ has
units of (shot-Torr)$^{-1}$ and depends on ion species. References to $%
\kappa N\left( I_{0}\right) $ refer to Guo's ordinate data itself, curve
fits to it, or theoretical values, all in the same units as Guo's plots, for
clarity. SI units are used otherwise. The background neutral gas density $%
n_{0}$ is sufficiently low that the $I$ profile is undisturbed by it and the
resulting ionization products, implying that $N\left( I_{0}\right) $ is
proportional to $n_{0}$. We, therefore, calibrate the $\kappa N\left(
I_{0}\right) $ data to determine the ratio $N\left( I_{0}\right) /n_{0}=$ $%
\kappa N\left( I_{0}\right) /\left( \kappa n_{0}\right) $, where $\kappa
n_{0}$ is species-dependent, but independent of $n_{0}$.

A correction factor $\beta $ to Guo's published $I_{0}$ scale and
calibration factor $\kappa n_{0}$ for Ar$^{+}$ are determined by adjusting
these parameters to fit the algorithm's $\left\langle W\right\rangle
_{\gamma }$ vs.\ $I$ solution to the ionization rate model of Peremlomov,
Popov, and Terent'ev (PPT) \cite{Perelomov66}\cite{Perelomov67}, which is
considered fairly accurate for noble gases \cite{Chin04}. The value of $%
\kappa n_{0}$ needed to invert the other species' data is then determined
from their relative MCP\ collection efficiencies, based on the
phenomenological model of Meier and Eberhart \cite{Meier93}. Meier does not
include results for Xe$^{+}$, so it is not considered, despite Guo
presenting count data for it.

\begin{figure}[H]\centering\includegraphics[width=3.375in]{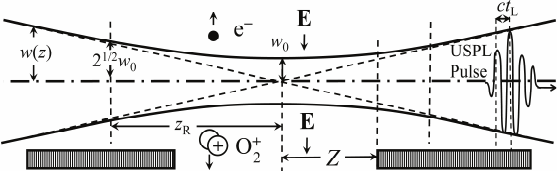}\end{figure} 
\textbf{Fig.~1 }Optical geometry of the aperture-limited focused Gaussian
USPL beam representing Guo's experiment.

\bigskip

The optical profile assumed and illustrated in Fig.~1 is a
diffraction-limited focused $\lambda =800$ nm ultrashort pulsed laser (USPL)
with both a Gaussian radial and temporal dependence $I=I\left( r,t\right) $.
Initially, a first order approximation to $\left\langle W\right\rangle
_{\gamma }$ vs.\ $I$ is presented that is a generalization of the method
used by Sharma, et al.\ \cite{Sharma18} to determine $\left\langle
W\right\rangle _{\gamma }$ vs.\ $I$ for O$_{2}$ from a direct measurement of 
$N\left( I_{0}\right) $. Sharma \cite{Sharma18} also assumes such an $I$
profile, but confines itself to the multiphoton regime, where $\left\langle
W\right\rangle _{\gamma }=\sigma _{8}I^{8}$ is a good approximation, and
ionization is weak (allowing the neutral density to be assumed constant).
For higher values of $I_{0}$, $\left\langle W\right\rangle _{\gamma }$ has a
more general relationship to $I$, but a power law with a lower real
coefficient $\mu $ may be assumed for $I\sim I_{0}$, provided $I_{0}$ is not
too high. Beyond \emph{this}, the first order approximation is still useful
as the starting point for a general inversion that makes no assumptions
about the form of $\left\langle W\right\rangle _{\gamma }$ vs.\ $I$, and
where neutral depletion \emph{is} significant.

Being based on ionization products subsequent to optical pulse passage, $%
\left\langle W\right\rangle _{\gamma }$ does not necessarily represent the
intrinsic ionization rate, but its effective value after passage. Electrons
which are released, but return to their parent ion within an optical cycle
of the oscillating $\mathbf{E}$-field and recombine are not counted. We
assume that such a population of recombined neutrals, likely in an excited
state and/or in the process of dissociation \cite{Talebpour96}, is either
small, or has a subsequent reionization rate of similar value. Guo, at
least, argues that the former is the case. Depletion of the majority ion
species by processes resulting in the other observed species reported on in
Guo is also not considered.

Details of the inversion method are presented first, but illustrated by Ar$%
^{+}$ results from data which has been calibrated by the PPT fit
subsequently described in Sec.~V. This fit, and its effect on the inversion
of all species, is parameterized by the values of correction factors to two
published properties. In addition to $\beta $ (defined above), $\alpha $ is
a correction factor to Sharma's published value of $\sigma _{8}$.\cite%
{Sharma18} $\alpha $ is independently estimated in the following to validate
the results presented, based on a reevaluation of Sharma's own data \cite%
{Sharma18}.

\section{Revision of Sharma's O$_{2}$'s multiphoton cross section $\protect%
\sigma_{8}$ for 800 nm radiation}

Sharma finds $\sigma _{8}=\sigma _{8,\mathrm{Sharma}}=3.3\times 10^{-130}$ W$%
^{-8}$m$^{16}$s$^{-1}$ \cite{Sharma18}, based on \emph{elastic} collision
frequency $\nu _{\mathrm{c,Sharma}}=5.18\times 10^{11}$ s$^{-1}$ ($\nu _{%
\mathrm{eg}}$ in Sharma \cite{Sharma18}), to which $\sigma _{8}$ is directly
proportional in its analysis.\ The propagation medium for the measurement is
air with a reported O$_{2}$ density of $n_{\mathrm{{O}_{2}}}=5.13\times
10^{24\mathrm{\ }}$m$^{-3}$. O$_{2}$'s number density is 21\% of air, so
this implies a total molecular density of $n_{\mathrm{air}}=2.44\times
10^{25}$ m$^{-3}$. This is less than standard atmospheric density $n_{%
\mathrm{L}}=2.6868\times 10^{25\mathrm{\ }}$m$^{-3}$ (Loschmidt's number 
\cite{NRL}), consistent with an altitude of roughly 800 m. The use of a test
cell with an adjustable background pressure, though, is reported in a
subsequent publication by this group \cite{Shashurin23}.

The electron temperature Sharma assumes is $T=0.4$ eV \cite{Sharma18}.
However, $2/3$ of the electron kinetic energy spectrum's mean kinetic energy
recorded for electrons traveling in the $\mathbf{E}$-field direction due to O%
$_{2}\ $(the main source of electrons in air at the low energies entailed 
\cite{Guo98}) exposed to a linearly polarized $\lambda =800$ nm USPL at $%
I_{0}=25.23$ TW/cm$^{2}$ (very near Sharma's $I_{0}=26.8$ TW/cm$^{2}$) is $%
T=0.65$ eV, based on integration of the spectrum of K{\l }oda, et al. \ \cite%
{Kloda10}. Sharma's actual $T$ \cite{Sharma18} is likely higher since its $%
I_{0}$ is higher, K{\l }oda's spectral rise appears to be the start of a
broad peak beyond the recorded range, and K{\l }oda only records electrons
traveling in the $\mathbf{E}$-field direction. Re the latter, there may be a
population of electrons that rescatter off their parent ion to other angles
that are subsequently accelerated to higher energies than would\ otherwise
be obtained \cite{Becker18}\cite{Okunishi08}. The \emph{momentum transfer}
collision frequency (relevant to diffusive transport) at the $T=0.65$ eV and 
$n_{\mathrm{air}}=2.44\times 10^{25}$ m$^{-3}$ is $\nu _{\mathrm{c}%
}=\allowbreak 1.59\times 10^{12}$ s$^{-1}$ \cite{Pusateri15}. This implies a
correction factor of at least $\alpha =\nu _{\mathrm{c}}/\nu _{\mathrm{%
c,Sharma}}=\allowbreak 3.06$ is needed for $\sigma _{8,\mathrm{Sharma}}$. We
have, then,%
\begin{equation}
\sigma _{8}>1.01\times 10^{-129}\mathrm{{\ W}^{-8}{m}^{16}{s}^{-1}}
\label{1}
\end{equation}%
The subject inversion of Guo's $\kappa N\left( I_{0}\right) $ data provides
an independent estimate of $\sigma _{8}$ and, therefore, Sharma's $T$ \cite%
{Sharma18}, given $\sigma _{8}$'s proportionality to $\nu _{\mathrm{c}}$ and
the relationship between $\nu _{\mathrm{c}}/n_{\mathrm{air}}$ and $T$ \cite%
{Pusateri15}.

\section{First order approximation of ionization rate $\left\langle
W\right\rangle _{\protect\gamma}$ vs.\ intensity $I$}

\ We generalize here Sharma's assumed power law relationship \cite{Sharma18}
between $\left\langle W\right\rangle _{\gamma }$ and $I$ for arbitrary
exponent $\mu $ (which need not be an integer), and assume the collection
volume is aperture-limited in the axial direction by a slit of half-width $Z$
centered on the laser's focal plane. This assumes the majority of electrons
are released during a time interval when the instantaneous local intensity $%
I $ of a pulse with a given peak intensity $I_{0}$ is within a range of
values for which $\left\langle W\right\rangle _{\gamma }=\sigma _{\mu
}I^{\mu }$ is accurate, where (unlike Sharma \cite{Sharma18}) $\mu $ and $%
\sigma _{\mu }$ are functions of $I_{0}$. Like Sharma \cite{Sharma18}, we
assume a focused laser beam with a Gaussian radial $I$ profile with $e^{-2}$
intensity half-width $w_{0}$, a Gaussian temporal $I$ waveform with $e^{-1}$
half-width $t_{\mathrm{L}}$, Rayleigh range $z_{\mathrm{R}}$ \cite{Siegman86}%
, and negligible depletion of the neutral molecules. We have, then,

\begin{equation}
\begin{tabular}{l}
$I=I_{0}\frac{w_{0}^{2}}{w^{2}\left( z\right) }\exp \left( -\frac{2r^{2}}{%
w^{2}\left( z\right) }-\frac{\left( t-t^{\ast }\right) ^{2}}{t_{\mathrm{L}%
}^{2}}\right) $ \\ 
$w^{2}\left( z\right) =w_{0}^{2}\left( 1+\left( \frac{z}{z_{\mathrm{R}}}%
\right) ^{2}\right) $ \\ 
$t^{\ast }=t^{\ast }\left( r,z\right) $ \\ 
$\left\langle W\right\rangle _{\gamma }=\sigma _{\mu }I^{\mu }$ \\ 
$n=n_{0}\sigma _{\mu }\int\nolimits_{-\infty }^{+\infty }I^{\mu }\left(
r,z,t\right) dt$ \\ 
$N\left( I_{0}\right) =\int\nolimits_{-Z}^{+Z}\int\nolimits_{0}^{+\infty
}2\pi rndrdz$%
\end{tabular}
\label{2}
\end{equation}

Upon solving the integral over $t$, 
\begin{equation}
\begin{tabular}{l}
$n=\frac{\pi ^{1/2}\kappa n_{0}t_{\mathrm{L}}}{\mu ^{1/2}\left( 1+\left( 
\frac{z}{z_{\mathrm{R}}}\right) ^{2}\right) ^{\mu }}\exp \left( -\frac{2\mu
r^{2}}{w_{0}^{2}\left( 1+\left( \frac{z}{z_{\mathrm{R}}}\right) ^{2}\right) }%
\right) $ \\ 
$\times \sigma _{\mu }I_{0}^{\mu }$%
\end{tabular}%
\ \   \label{3}
\end{equation}%
Substituting this into Eq.~\ref{2}.6 (referring here to the equation which
starts on the 6'th line of Eqs.~\ref{2}), solving the integral over $r$, and
changing the axial integral's integration variable from $z$ to $s=\left(
z/z_{\mathrm{R}}\right) ^{2}$,%
\begin{equation}
\begin{tabular}{l}
$N\left( I_{0}\right) =\left( \pi ^{3/2}\lambda n_{0}w_{0}^{2}\right) t_{%
\mathrm{L}}A_{\mu }\left( S\right) \sigma _{\mu }I_{0}^{\mu }$\ \ \ \  \\ 
$A_{\mu }\left( S\right) =\frac{z_{\mathrm{R}}}{2\mu ^{3/2}\lambda }%
\int\nolimits_{0}^{S}s^{-1/2}\left( 1+s\right) ^{1-\mu }ds$ \ \ 
\end{tabular}
\label{4}
\end{equation}%
where $S=\left( Z/z_{\mathrm{R}}\right) ^{2}$. The integral over $s$ is
equal to $2\sqrt{S}F\left( \mu -1,\frac{1}{2};\frac{3}{2};-S\right) $, where 
$F=$ $_{2}F_{1}$ \ is the hypergeometric function, based on Gradshteyn and
Ryzhik \cite{Gradshteyn15} (\textquotedblleft G\&R\textquotedblright ) Eq.
3.194.1. $F$ is initially expressed as G\&R Eq.~9.111, but analytically
continued for $S>1$ per G\&R Sec.~9.154. We solve for the limiting cases of $%
S\rightarrow \infty $ and $S\rightarrow 0$ separately, resulting in,

\begin{equation}
\begin{tabular}{l}
$A_{\mu}\left( S\right) =\frac{z_{\mathrm{R}}\sqrt{S}}{\mu^{3/2}\lambda }%
F\left( \mu-1,\frac{1}{2};\frac{3}{2};-S\right) $ \\ 
$A_{\mu}\left( \infty\right) =\frac{8\pi z_{\mathrm{R}}\Gamma\left(
2\mu-3\right) }{\mu^{3/2}\lambda\left( 2^{\mu}\Gamma\left( \mu-1\right)
\right) ^{2}}$ \\ 
$A_{\mu}\left( 0\right) =\frac{Z}{\mu^{3/2}\lambda}$%
\end{tabular}
\label{5}
\end{equation}

$S\rightarrow \infty $ corresponds to an unlimited aperture (such as with
Sharma \cite{Sharma18}), for which the integral in Eq.~\ref{4}.2 is $B\left(
1/2,\mu -3/2\right) $ $=$ $\Gamma \left( 1/2\right) \Gamma \left( \mu
-3/2\right) /\Gamma \left( \mu -1\right) $ (G\&R Eq.~8.384.1), where $%
B\left( x,y\right) $ and $\Gamma \left( x\right) $ are the beta and gamma
functions, respectively. To eliminate the half-integer terms, we use for Eq.~%
\ref{5}.2 $\Gamma \left( 1/2\right) =\sqrt{\allowbreak \pi }$ from G\&R
Eq.~8.338.2, and G\&R Eq.~8.335.1 with argument $x=\left( \mu -3/2\right) $
to derive $\Gamma \left( 1/2\right) \Gamma \left( \mu -3/2\right) /\Gamma
\left( \mu -1\right) =16\pi \Gamma \left( 2\mu -3\right) \left( 2^{2\mu
}\Gamma \left( \mu -1\right) \right) ^{-2}$. The result (Eq.~\ref{5}.2) is
equivalent to Sharma's Eq.~4 \cite{Sharma18} for $\mu =8$.

$S\rightarrow 0$, meanwhile, corresponds to $Z$ being sufficiently small
relative to $z_{\mathrm{R}}$ that we may assume that the laser beam profile
sampled is axially invariant. To determine $A_{\mu }\left( S\right) $, then,
we do \emph{not} change the integration variable to $s=\left( z/z_{\mathrm{R}%
}\right) ^{2}$ in Eq.~\ref{3} for the integration over $z$ in Eq.~\ref{2}.6
to determine $N\left( I_{0}\right) $, but instead take $z_{\mathrm{R}%
}\rightarrow \infty $ in Eq.~\ref{3}. Eq.~\ref{5}.3, then, follows from Eq.~%
\ref{2}.6 and Eq.~\ref{4}. This approximation should be used with caution,
since $Z$ may have to be over an order of magnitude smaller than $z_{\mathrm{%
R}}$ for reasonable accuracy in cases where the dependence of $\left\langle
W\right\rangle _{\gamma }$ on $I$ is strong.

To estimate $S$ for inversion of Guo's $\kappa N\left( I_{0}\right) $ data 
\cite{Guo98}, we take $Z=0.5$ mm, which is the radius of the pinhole used to
aperture-limit ion transport for counting. The value of $z_{\mathrm{R}}$ is
not provided, but may be estimated with sufficient accuracy for our purposes
(as will be shown) from the relationship $z_{\mathrm{R}}=\pi
w_{0}^{2}/\lambda $ for a focused Gaussian beam \cite{Siegman86}, where $%
w_{0}$ is the value for which Guos' maximum beam pulse energy of $U=400$ $%
\mu $J results in the maximum value of $I_{0}=I_{0,\max }\approx 7.94\times
10^{14}$ W/cm$^{2}$ for their plotted results. Note, though, that the
precise value of $I_{0,\max }$ depends on a recalibration of Guo's $I_{0}$
scale. Integration of Eq.~\ref{2}.1 over space-time results in the
relationship $U=\pi ^{3/2}I_{0,\max }t_{\mathrm{L}}w_{0}^{2}/2$, so,%
\begin{equation}
w_{0}=\sqrt{\frac{2U}{\pi ^{3/2}I_{0,\max }t_{\mathrm{L}}}}\mathrm{\ \ \ \ \
\ \ \ }z_{\mathrm{R}}=\frac{\pi w_{0}^{2}}{\lambda }\   \label{6}
\end{equation}

The USPL $e^{-1}$ half-width $t_{\mathrm{L}}=18$ fs is determined from their
laser pulse width of $30$ fs, which we take to be the full width at half max
(based on a description of the laser in another article \cite{Li98}) divided
by $2\sqrt{\ln \left( 2\right) }$. $S$ is sufficiently small that the $S=0$
approximation of Eq.~\ref{5}.3 is a reasonable approximation. This implies
the accuracy of our $z_{\mathrm{R}}$ estimate is not critical. We will,
nonetheless, include the general form of $A_{\mu }\left( S\right) $ in our
analysis so that it may be applied to cases where it \emph{is} significant.

\begin{figure}[H]\centering\includegraphics[width=3.375in]{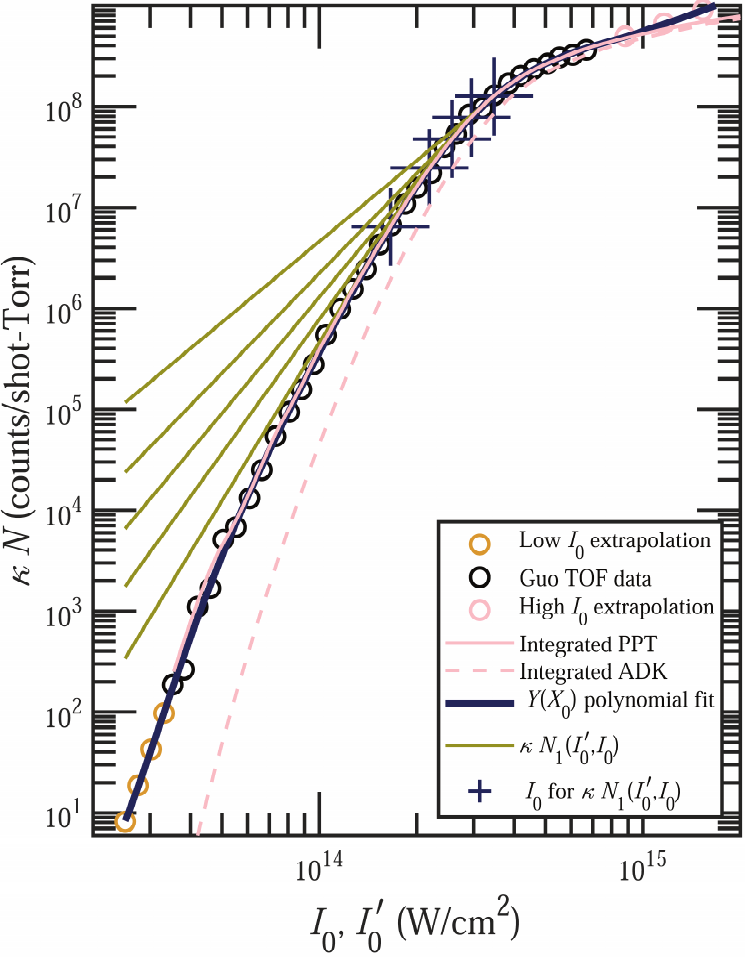}\end{figure}  
\textbf{Fig.~2} A $\kappa N\left( I_{0}\right) $ plot of Guo's Ar$^{+}$ data
(dark circles), extrapolated pseudo-data (lighter circles), and the fit to
both (black line) are overlaid onto the space-time integral of the PPT model
(Sec.~V) that the inverted $\left\langle W\right\rangle _{\gamma }$ vs.\ $I$
solution is fit to. The corresponding ADK model (Sec.~V) is shown for
reference. Crosses show where neutral Ar is depleted at maximum focus by
approximately 10, 30, 50, 70, and 90 percent (Sec.~IV). Tangent lines
depicting the first order approximation $\kappa N_{1}\left( I_{0}^{\prime
},I_{0}\right) $ vs.\ $I_{0}^{\prime }$ at these $I_{0}$ values are also
plotted.

\bigskip

To apply our first order approximation to Guo's log-log plots of $\kappa
N\left( I_{0}\right) $, we approximate $\kappa N\left( I_{0}^{\prime
}\right) $ for a range of peak intensities $I_{0}^{\prime }$ $\lesssim $ $%
I_{0}$ for any given $I_{0}$ by $\kappa N_{1}\left( I_{0}^{\prime
},I_{0}\right) $, where $N_{1}\left( I_{0}^{\prime },I_{0}\right) $ has the
form of the power law Eq.~\ref{4}.1 in terms of $I_{0}^{\prime }$, with $\mu 
$ and $\sigma _{\mu }$ being functions of $I_{0}$ alone. \ We assign, by
directly fitting to Guo's plots,%
\begin{equation}
\begin{tabular}{l}
$X_{0}=\log \left( I_{0}\right) $ \ \ $\ Y\left( X_{0}\right) =\log \left(
\kappa N\left( I_{0}\right) \right) $ \\ 
$X_{0}^{\prime }=\log \left( I_{0}^{\prime }\right) $ \ $\ \ Y_{1}\left(
X_{0}^{\prime },X_{0}\right) =\log \left( \kappa N_{1}\left( I_{0}^{\prime
},I_{0}\right) \right) $%
\end{tabular}
\label{9}
\end{equation}%
where, like Guo, the logs are base 10. From Eq.~\ref{4}.1, then,%
\begin{equation}
\begin{tabular}{l}
$\kappa N_{1}\left( I_{0}^{\prime },I_{0}\right) =KA_{\mu }\left( S\right)
\sigma _{\mu }I_{0}^{\prime \mu }$ \\ 
$Y_{1}\left( X_{0}^{\prime },X_{0}\right) =\log \left( KA_{\mu }\left(
S\right) \sigma _{\mu }\right) +\mu X_{0}^{\prime }$ \\ 
$\ K=\pi ^{3/2}\lambda w_{0}^{2}t_{\mathrm{L}}\kappa n_{0}$%
\end{tabular}
\label{10}
\end{equation}

One sees from Eq.~\ref{10}.2, and as plotted in Fig.~2 for select values of $%
I_{0}$, that $\left\langle W\right\rangle _{\gamma }=\sigma _{\mu }I^{\prime
\mu }$ is a good approximation for values of $I_{0}$ below where neutral
depletion is $\leq 10\%$, since $Y\left( X_{0}\right) $ can be approximated
by a straight line of slope $\mu $ over a significant range of $%
I_{0}^{\prime }\leq I_{0}$ then. This line is the tangent at $%
X_{0}=X_{0}^{\prime }$. The least-squares fit of the data points plotted
(circles) to an $8$-order polynomial in (offset) variable $X_{0}-X_{1}$ is
seen in Fig.~2 to provide an accurate approximation $Y_{\mathrm{fit}}\left(
X_{0}\right) $ to $Y\left( X_{0}\right) $. It is used to determine $\mu $
vs.\ $I_{0}$, 
\begin{equation}
\begin{tabular}{l}
$Y_{\mathrm{fit}}\left( X_{0}\right) =\sum\limits_{i=0}^{m}a_{i}\left(
X_{0}-X_{1}\right) ^{i}$ \\ 
$\mu =D_{X_{0}}Y_{\mathrm{fit}}\left( X_{0}\right)
=\sum\limits_{i=1}^{m}a_{i}i\left( X_{0}-X_{1}\right) ^{i-1}$ \\ 
$X_{1}=\log \left( 100\mathrm{{\ TW}/{cm}^{2}}\right) $%
\end{tabular}%
\ \ \ \   \label{12}
\end{equation}%
where $D_{x}$ is the derivative operator w.r.t.\ $x$. The crosses in Fig.~2
(and subsequent figures) and tangent first order approximations (straight
lines)\ are placed at values of $I_{0}$ for which the numerical solution to $%
n/n_{0}$ peaks on the computational mesh of the general inversion algorithm
(presented in the next section) closest to, but below, 0.1, 0.3, ... , 0.9.
The high level of depletion beyond this accounts for $\kappa N\left(
I_{0}\right) $ increasing much more slowly with $I_{0}$ there.

Artificial data points (light circles in Fig.~2) linearly extrapolate Guo's $%
Y$ vs.\ $X_{0}$ data (black circles) both below and above the actual data
range. They are included with the data that the polynomial of Eq.~\ref{12}.1
is fit to. The purpose of the upper extrapolation is purely numerical; it is
found to permit the general inversion to achieve a verified solution for
higher values of $I_{0}$ than it otherwise would, based on the standard
defined in Sec.~IV. The lower extrapolation exploits the fact that the first
order approximation to $\left\langle W\right\rangle _{\gamma }$ is accurate
for such low values of $I$, and fulfills the general inversion's need to
consider $\left\langle W\right\rangle _{\gamma }$ off-peak (where $I\mathcal{%
<}I_{0}$). This enables the algorithm to provide a solution to $\left\langle
W\right\rangle _{\gamma }$ all the way down to $I\mathcal{=}I_{0}$ of Guo's
lowest \emph{actual} $\kappa N\left( I_{0}\right) $ data. Solving Eq.~\ref%
{10}.1 at $I_{0}^{\prime }=I_{0}$ (where $\kappa N_{1}\left( I_{0}^{\prime
},I_{0}\right) =\kappa N\left( I_{0}\right) $) for $K$, we have at the
lowest $I_{0}$ for which there is actual data,

\begin{equation}
K=\frac{\kappa N\left( I_{00}\right) }{A_{\mu _{00}}\left( S\right) \sigma
_{\mu _{00}}I_{00}^{\mu _{00}}}  \label{14}
\end{equation}%
$I_{00}$ and $\mu _{00}$ are the values of $I_{0}$ and $\mu $ at this data's 
$I_{0}$. The solution to $\sigma _{\mu _{00}}$ is a subject of Sec.~V.

Once $K$ is determined for a given species, then $\sigma_{\mu}$ elsewhere is
determined by solving Eq.~\ref{10}.1 for it at $I_{0}=I_{0}^{\prime}$
(again, where $\kappa N_{1}\left( I_{0}^{\prime},I_{0}\right) =\kappa
N\left( I_{0}\right) $). From this and Eq.~\ref{2}.4, our first order
solution is specified by, 
\begin{equation}
\sigma_{\mu}=\frac{\kappa N\left( I_{0}\right) }{KA_{\mu}\left( S\right)
I_{0}^{\mu}}\mathrm{\ \ \ \ \ \ \ }\left\langle W\right\rangle
_{\gamma}=\sigma_{\mu}I^{\mu}\mathrm{\ }  \label{15}
\end{equation}

\section{General inversion}

The first order approximation to $\left\langle W\right\rangle _{\gamma }$
vs.\ $I$ is used as the starting point for an iterative solution to a more
general inversion that assumes $D_{t}n\left( t\right) =\left( n_{0}-n\left(
t\right) \right) \left\langle W\right\rangle _{\gamma }$. $\ n\left(
t\right) $, here, is the time dependent ion density distribution during the
pulse. $n$ (as previously used, with no parenthetical) is $n\left( \infty
\right) $ (after optical pulse passage). Unlike the first order
approximation itself, the prefactor $\left( n_{0}-n\left( t\right) \right) $
accounts for neutral depletion reducing the ionization rate density due to
the main ionization path. Although data for other ion species is presented
in Guo, taking them into account entails a series of coupled differential
equations relating the ionization rates of the various parents and
daughters. These, fortunately, have a relatively minor contribution ($\sim
10^{-2}$) in the $I$ range for which our algorithm may be applied ($<90\%$
peak neutral depletion ), so are not considered. The solution is,%
\begin{equation}
\frac{n}{n_{0}}=1-\exp \left( -\int\nolimits_{-\infty }^{+\infty
}\left\langle W\right\rangle _{\gamma }dt\right)  \label{16}
\end{equation}

\begin{figure}[H]\centering\includegraphics[width=3.375in]{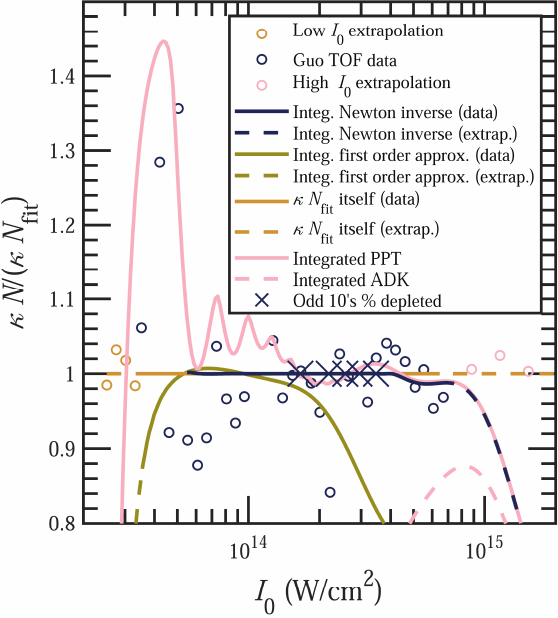}\end{figure}   
\textbf{Fig.~3 }This plot shows the ratio of various $\kappa N\left(
I_{0}\right) $ results (listed in the legend) for Ar$^{+}$ to the 8-order
fit to Guo's data $\kappa N_{\mathrm{fit}}\left( I_{0}\right) $. The solid
black line is this ratio for $\kappa N_{\mathrm{R}}\left( I_{0}\right) $, as
reconstructed from the data inversion's final $\left\langle W\right\rangle
_{\gamma }$ solution. The circles are such ratios for Guo's data (black) and
extrapolated pseudodata (light) to which $\kappa N_{\mathrm{fit}}\left(
I_{0}\right) $ is fit. Ratios for the PPT and ADK (Sec.~V) models'
space-time integrals, crosses near the odd 10's of percent neutral
depletion, \ and $\kappa N_{\mathrm{R}}\left( I_{0}\right) $ for the first
order approximation to $\left\langle W\right\rangle _{\gamma }$ are also
plotted.

\bigskip

To initialize the general inversion, the integral of $\left\langle
W\right\rangle _{\gamma }$ over $t$ of the first order approximation (Eqs.~%
\ref{2}) is solved by doubling its integral over\ the interval $-3t_{\mathrm{%
L}}\leq \left( t-t^{\ast }\right) \leq 0$ at finely spaced intervals of $r$, 
$z$, and $I_{0}$ within the ranges $0$ $\leq r\leq 2w\left( Z\right) $, $%
0\leq $ $z\leq Z$, and $I_{0,\min }\leq I_{0}\leq I_{0,\max }$, where the
range of $I_{0}$ here includes the extrapolated pseudo-data. Twice the
integral of $n/n_{0}$ over the specified half-volume, then, provides a
reconstruction of $N\left( I_{0}\right) /n_{0}$ for each $I_{0}$, based on
the first order approximation.

The reconstructed $N\left( I_{0}\right) /n_{0}$ so obtained is then
multipled by $\kappa n_{0}$ to obtain first order measurement reconstruction 
$\kappa N_{\mathrm{R}}\left( I_{0}\right) $ where, from Eq.~\ref{10}.3,%
\begin{equation}
\kappa n_{0}=\frac{K}{\pi ^{3/2}\lambda w_{0}^{2}t_{\mathrm{L}}}\ \ 
\label{17}
\end{equation}%
$K$ here is found from Eq.~\ref{14}, given $\sigma _{\mu _{00}}$ from
Sec.~V. $\kappa N_{\mathrm{fit}}\left( I_{0}\right) $ of Guo's measurements,
based on the logorithmic form of Eq.~\ref{12}.1, is formally compared to the
fit $\kappa N_{\mathrm{R}}\left( I_{0}\right) $ in the following.

\begin{figure}[H]\centering\includegraphics[width=3.375in]{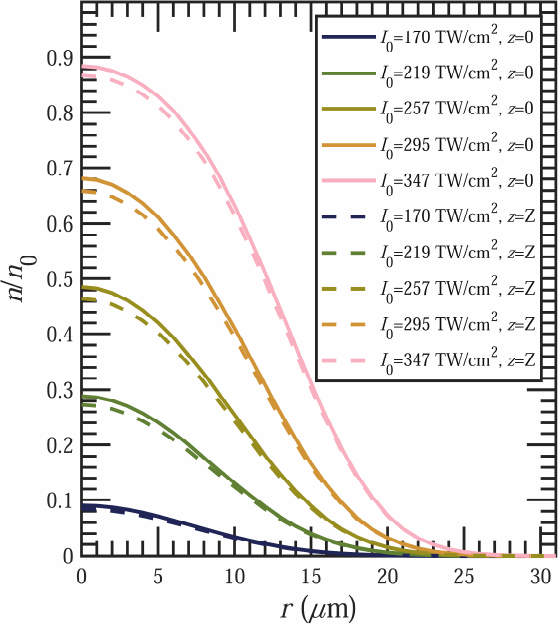}\end{figure}  
\textbf{Fig.~4. }$n/n_{0}$ vs.\ $r$ at center ($z=0)$ and edge of ($z=Z$) of
the aperture are plotted for the values of $I_{0}$ shown in the legend,
corresponding to peak neutral Ar depletion being near the odd 10's of
percent.

\bigskip

A method generalizing Newton's method to zero a function of a single
variable $x$ is used to find the solution to $\left\langle W\right\rangle
_{\gamma }$\ vs.\ $I$ consistent with Guo's data. The method, though, is
generalized to zero the \emph{functional} $f\left[ x\right] =\kappa N_{%
\mathrm{fit}}\left( I_{0}\right) -\kappa N_{\mathrm{R}}\left( I_{0}\right) $%
, where $x=\left\langle W\right\rangle _{\gamma }$ vs. (local)\ $I$. $\kappa
N_{\mathrm{R}}\left( I_{0}\right) $ is the reconstructed signal, calculated
as done for the first order solution above, but for each iteration's
increasingly refined estimate of $\left\langle W\right\rangle _{\gamma }$
vs.\ $I$. $f\left[ x\right] $ is a functional (function of a function) in
the sense that $x=\left\langle W\right\rangle _{\gamma }$ is itself a
function of $I$. By this method, iteration $i+1$ of $x$ is, 
\begin{equation}
\begin{tabular}{l}
$x_{i+1}=x_{i}-\left\langle \frac{f\left[ x_{i}\right] }{f^{\prime }\left[
x_{i}\right] }\right\rangle _{N_{\mathrm{box}}}$ \\ 
$f\left[ x\right] =\kappa N_{\mathrm{fit}}\left( I_{0}\right) -\kappa N_{%
\mathrm{R}}\left( I_{0}\right) $ \\ 
$f^{\prime }\left[ x\right] =\frac{f\left[ x+\delta x\right] -f\left[ x%
\right] }{\delta x}$ \\ 
$x=$ $\left\langle W\right\rangle _{\gamma }$ \ \ \ $\delta
x=10^{-4}\left\langle W\right\rangle _{\gamma }$%
\end{tabular}%
\ \ \   \label{18}
\end{equation}%
$x_{0}$ is the first order solution to $\left\langle W\right\rangle _{\gamma
}$ vs.\ $I$. $\left\langle {}\right\rangle _{N_{\mathrm{box}}}$ is an
operation that the original Newton's method does \emph{not} employ. It
refers to the data representing the enclosed functional's value vs.\ $I_{0}$
being smoothed by (boxcar) averaging the $N_{\mathrm{box}}$ nearest
neighbors (including itself) for each value of $I_{0}$ it is calculated for. 
$f\left[ x\right] $ is calculated at equal intervals of $0.01$ of $I_{0}$'s
logorithmic form $X_{0}=\log \left( I_{0}\mathrm{{\ }\left[ {W}/{m}^{2}%
\right] }\right) $. $\ N_{\mathrm{box}}=21$, $31$, and $15$ for Ar$^{+}$, O$%
_{2}^{+}$, and N$_{2}^{+}$, respectively.

Boxcar averaging is necessary for $f\left[ x\right] $ to achieve a value
near zero. The ratio of $\kappa N_{\mathrm{R}}\left( I_{0}\right) $ to $%
\kappa N_{\mathrm{fit}}\left( I_{0}\right) $, after several iterations, is
plotted in Fig.~3 for Ar$^{+}$. Its deviation from unity by less than 0.1\%
is the standard for verifying the range of $I_{0}$ for which the inversion
provides a mathematically accurate solution, despite (as will be shown)
unphysical behavior beyond a value of $I_{0}$ within this formally verified
range.

As plotted in Fig.~3, The generalized Newton's method is only used to
determine $\left\langle W\right\rangle _{\gamma }$ vs.\ $I$ for $I=I_{0}$
greater than the value for which first order $\kappa N_{\mathrm{R}}\left(
I_{0}\right) /\left( \kappa N_{\mathrm{fit}}\left( I_{0}\right) \right) $
first approaches unity. Below this $I_{0}$,$\ \left\langle W\right\rangle
_{\gamma }=\sigma _{\mu }I^{\mu }$ itself (from Eqs.~\ref{15} and Eq.~\ref%
{12}.2) is used for each iteration \emph{and} the final result. This is
necessary since a range of $I$ significantly less than $I_{0}$ is needed to
perform the numerical space-time integral to determine $\kappa N_{\mathrm{R}%
}\left( I_{0}\right) $. This procedure ensures a continuous smooth
transition between the first order approximation (in a range where it is
accurate) and the generalized Newton method solution beyond.

Figure~4 plots $n/n_{0}$ vs.\ $r$ at $z=0$ and $z=Z$, based on Eq.~\ref{16},
for Ar$^{+}$ for\ a range of $I_{0}$ values for which its solution peaks on
the computational mesh closest to, but less than, 0.1, 0.3, ... , 0.9, as
marked in the prior figures. It is clear from this that the large $z_{%
\mathrm{R}}$ approximation $A_{\mu }\left( S\right) \approx A_{\mu }\left(
0\right) $ would have been appropriate for this geometry, and that,
therefore, the accuracy of our $w_{0}$ estimate is not critical.

\section{Calibrating Guo's $I_{0}$ and $\protect\kappa N\left( I_{0}\right) $
scales}

The ratio of $\kappa N_{1}\left( I_{00},I_{00}\right) $ for any two species
is, from Eq.~\ref{10}.1, Eq.~\ref{10}.3, and Eq.~\ref{5}.1,%
\begin{equation}
\begin{tabular}{l}
$\frac{\kappa _{i}N_{i}\left( I_{00i}\right) }{\kappa _{j}N_{j}\left(
I_{00j}\right) }=\frac{\mu _{\mathrm{{r}i}}}{\mu _{\mathrm{{r}j}}}\frac{\mu
_{00j}^{3/2}\sigma _{\mu _{_{00}}i}I_{00i}^{\mu _{00i}}}{\mu
_{00i}^{3/2}\sigma _{\mu _{00}j}I_{00j}^{\mu _{00j}}}\frac{F\left( \mu
_{00i}-1,\frac{1}{2};\frac{3}{2};-S\right) }{F\left( \mu _{00j}-1,\frac{1}{2}%
;\frac{3}{2};-S\right) }$ \\ 
$\frac{\mu _{\mathrm{{r}i}}}{\mu _{\mathrm{{r}j}}}=\frac{m_{i}a_{i}v_{i}%
\arctan \left( b_{i}\left( v_{i}-v_{0}\right) \right) }{m_{j}a_{j}v_{j}%
\arctan \left( b_{j}\left( v_{j}-v_{0}\right) \right) }$ \ \ $v=\sqrt{\frac{%
2V_{\mathrm{a}}K_{\mathrm{eV}}}{Mm_{\mathrm{amu}}}}$%
\end{tabular}%
\ \ \ \ \   \label{22}
\end{equation}%
where $i$ and $j$ represent their molecular symbols. Here and below, adding
a molecule's symbol or variable representation of it (as is the case here)
to a subscript refers to that molecule's singly charged ion value. $\kappa
_{i}/\kappa _{j}$ in the l.h.s.\ of Eq.~\ref{22}.1 is assumed equal to the
ratio of the relative collection efficiencies $\mu _{\mathrm{{r}i}}/\mu _{%
\mathrm{{r}j}}$ in the r.h.s., based on Eq.~3 of Meier and Eberhardt \cite%
{Meier93}. The $m$ terms are the number of atoms in each molecule, the $a$
and $b$ terms are phenomenological parameters associate with the molecules'
atomic constituents, $v$ is the ion impact velocity on the MCP, $V_{\mathrm{a%
}}$ is the net acceleration voltage determining impact kinetic energy, $K_{%
\mathrm{eV}}=1.6022\times 10^{-19}$ J/eV, $M$\ is the molecular mass in amu, 
$m_{\mathrm{amu}}=1.6605\times 10^{-27}$ kg/amu, and $v_{0}=3.3\times 10^{4}$
m/s is the cutoff speed for Meier's NiCr coating ($\mu _{\mathrm{r}}=0$ for $%
v\leq v_{0}$). $V_{\mathrm{a}}=3.1$ kV, based on personal correspondence\
with Guo's coauthor Prof. George N. Gibson. The $a$ and $b$ parameters are
tabulated in Meier's Table 2 for the elemental constituents of O$_{2}$ and N$%
_{2}$, and determined by a least squares fit to the data in its Table II for
Ar$^{+}$. We have, then,

\begin{center}
\textbf{Table 1}
\end{center}

\begin{equation*}
\begin{tabular}{l|l|ll}
& Ar$^{+}$ & O$_{2}^{+}$ & N$_{2}^{+}$ \\ \hline
$m$ & $1$ & $2$ & $2$ \\ 
$a/10^{-6}$ (s/m) & $8.73$ & $7.34$ & $4.26$ \\ 
$b/10^{-6}$ (s/m) & $16.7$ & $7.38$ & $18.8$ \\ 
$\mu_{\mathrm{r}}$ & $1.047$ & $1.311$ & $1.409$ \\ 
$v/10^{5}$ (m/s) & $1.367$ & $1.224$ & $1.461$%
\end{tabular}
\ \ \ 
\end{equation*}

Meier's data is based on an NiCr MCP electrode coating. Prof. Gibson states
they used \textquotedblleft standard Galileo plates from the mid
1990s\textquotedblright\ for the measurements. \ Ni-Cr based alloys, such as
80-20 Nichrome \cite{Popecki16} or Inconel\ \cite{Wiza79} (which have
similar percentages of Ni and Cr) are fairly standard for MCP's. The Galileo
group in 1994 \cite{Snider94} discusses their MCP\ fabrication techniques in
which Ni \emph{or} Cr is used. These are for experimental devices, however.
Alloying the two is usual for commercial electrical conductor applications
potentially subject to great heat due the superior mechanical properties
that result. Indeed, University of Nevada, Reno's Material Safety Data
Sheet, dated 1995, for \textquotedblleft Galileo Electro-Optics Corporation
Microchannel Plates and Single Channel Detectors\textquotedblright\ reports
both \textquotedblleft Nickel Oxide\textquotedblright\ and \textquotedblleft
Chromium Oxide\textquotedblright\ as \textquotedblleft potentially hazardous
components\textquotedblright . We, therefore, assume that Guo's MCP\ used a
NiCr alloy sufficiently similar to Meier to justify use of its collection
efficiency model. Further investigation is deemed unwarranted since Meier's
Table 3 records that the ratio of atomic yield using NiCr vs.\ Cu coating
only varies by an amount of order 10\% for atomic masses greater than that
of N. Given that the collection efficiency ratios for our three molecules of
interest do not stray far from unity, this would only result in a few to
several percentage error for $\left\langle W\right\rangle _{\gamma }$.

Given the ratio of collection efficiencies for the MCP of any two gas
species, $\sigma _{\mu _{00}}$ for one can be determined from the other by
solving Eq.~\ref{22}.1 for it. This allows us determine $K$ from Eq.~\ref{14}%
, and then $\kappa n_{0}$ from Eq.~\ref{17}, for both. $I_{00\mathrm{{O}_{2}}%
}$ is sufficiently small for $\sigma _{\mu _{00\mathrm{{O}_{2}}}}$ to
approximate O$_{2}$'s multiphoton cross section $\sigma _{8}$ \cite{Sharma18}%
. The correction factor $\alpha $, then, to Sharma's value \cite{Sharma18},
in addition to $\beta $, are sufficient to calculate $\kappa n_{0}$ and,
therefore, determine the solution to $\left\langle W\right\rangle _{\gamma }$
vs.\ $I$ for all treated species. For this reason, and to improve the
consistency with Sharma's data \cite{Sharma18} (which extends to lower
values of $I_{0}$), the low $I_{0}$ extrapolated pseudo-data for O$_{2}^{+}$%
's 8-order fit is determined by the line of slope $\mu _{00\mathrm{{O}_{2}}%
}=8$ that passes through the \emph{second} lowest value in Guo's $Y_{0}$
vs.\ $X_{0}$ actual data (the first point appears to be an outlier). Lacking
such data for Ar$^{+}$ and N$_{2}^{+}$, their extrapolated pseudo-data
points are determined by a line fit to their lowest $4$ actual data points.

Guo's original $I_{0}$ scale calibration is performed by adjusting the $%
I_{0} $ and $\kappa N\left( I_{0}\right) $ scales for the Ar$^{+}$ data to
fit the space-time integrated ionization rate density of the Ammosov,
Delone, and Kra\u{\i}nov (ADK) model \cite{Ammosov86} at their highest $%
I_{0} $ range. This is done since ADK is considered accurate for Ar $%
\rightarrow $ Ar$^{+}+$ e$^{-}$ for $\gamma <0.5$ \cite{Ilkov92}, where, for
linear polarized light, $\gamma =\left( \omega _{0}/e\right) \sqrt{c\epsilon
_{0}mU_{0}/I}$ is the Keldysh parameter \cite{Keldysh65}, $U_{0}$ is the
ionization potential from the ground state, $\omega _{0}=2.3545\times
10^{15} $ s$^{-1}$ is the laser's optical angular frequency, $e$ is electron
charge, $m$ is electron mass, and $\epsilon _{0}$ is the permittivity of
free space. Such accuracy may be the case, but Chin \cite{Chin04} notes that
fitting $\kappa N\left( I_{0}\right) $ data for Ar$^{+}$ to ADK for $\gamma
<0.5$ for the $200$ fs laser discussed in that reference is unreliable since
Ar neutrals are strongly depleted in this regime. Such depletion implies
that $\kappa N\left( I_{0}\right) $ is only representative of the depleted
volume which, in turn, depends on the accuracy to which the laser pulse
represents a Gaussian distribution in both space and time far off-peak \emph{%
for a particular laser system}. Sharma \cite{Sharma18} makes an effort to
assure that its $I$ distribution is accurately Gaussian in both space and
time, but such is not reported to be the case for either Chin or Guo. This
is not as big of an issue with Guo's (shorter, but still strongly depleting) 
$30$ fs pulse. Nonetheless, as shown below, it limits our ability to invert $%
\kappa N\left( I_{0}\right) $ vs.\ $I_{0}$ data above the value of $I_{0}$
for which peak neutral depletion exceeds 90\%.

We determine Guo's $I_{0}$ and $\kappa N\left( I_{0}\right) $ scale factors $%
\beta$ and (via $\alpha$) $\kappa n_{0\mathrm{Ar}}$, instead, by fitting Ar$%
^{+}$ data to the PPT\ model (described below), which fits well for noble
gases ions such as Ar$^{+}$ for much lower values of $I$ than ADK \cite%
{Chin04}. As a further refinement (enabled by the inversion technique
presented here), instead of fitting $\kappa N\left( I_{0}\right) $ to the
space-time integrated PPT model, we fit the \emph{inverted} $\left\langle
W\right\rangle _{\gamma}$ vs.\ $I$ curve to the \emph{unintegrated} PPT\
model. This results is a more accurate calibration, since it is much more
sensitive to small changes in the scale factors than $\kappa N\left(
I_{0}\right) $. The slope $\mu_{00}$ of the log-log plot of $\kappa N\left(
I_{0}\right) $ at $I\mathcal{=}I_{00}$, and measurement $\kappa N\left(
I_{00}\right) $ there are unaffected by changes to either $\beta$ or $\kappa
n_{0}$, although $I_{00}$ itself and $\sigma_{00}$ are.

The Ar$^{+}$ PPT results presented are based on Eq.~54 of Perelomov, et al.,
1966 \cite{Perelomov66} for Ar $\rightarrow $ Ar$^{+}+$e$^{-}$multiplied by
the long range Coulomb correction factor for linear polarization $2F_{0}/F$
of Eq.~45 of Perelomov, et al., 1967 \cite{Perelomov67}. \ $F_{0}$ and $F$
are the electric field magnitudes associated with the ground state electron
and the laser envelope, respectively, in atomic units. $F_{0}=\left(
2U_{0}/U_{\mathrm{au}}\right) ^{3/2}$ and $F=\sqrt{2I/\left( c\epsilon
_{0}\right) }/\mathcal{E}_{\mathrm{au}}$, where $U_{\mathrm{au}}=27.21$ eV
and $\mathcal{E}_{\mathrm{au}}=$ $5.141\times 10^{11}$ V/m are the atomic
units for energy and electric field, respectively \cite{Weinhold12}. The
result is,%
\begin{equation}
\begin{tabular}{l}
$w\left( \omega ,F\right) =\left( \frac{6}{\pi }\right) ^{1/2}\left\vert
C_{n^{\ast }l^{\ast }}^{\ }\right\vert ^{2}f\left( l,m\right) E_{0}$ \\ 
$\times $ $\left( 1+\gamma ^{2}\right) ^{\left\vert m\right\vert
/2+3/4}\left( \frac{2F_{0}}{F}\right) ^{2n^{\ast }-\left\vert m\right\vert
-3/2}$ \\ 
$\times $ $A_{m}\left( \omega ,\gamma \right) \exp \left( -\frac{2F_{0}}{3F}%
g\left( \gamma \right) \right) $%
\end{tabular}%
\ \ \ \   \label{20}
\end{equation}%
$n^{\ast }=\sqrt{U_{\mathrm{au}}/\left( 2U_{0}\right) }$ for emission from
the ground state, $E_{0}=U_{0}/U_{\mathrm{au}}$, $\omega =$ $\hbar \omega
_{0}/U_{\mathrm{au}}$, and $w\left( \omega ,F\right) =\left\langle
W\right\rangle _{\gamma }t_{\mathrm{au}}$, where $\hbar \omega _{0}$ is
laser photon energy in eV and $t_{\mathrm{au}}=2.4189\times 10^{-17}$ s is
the unit of time in atomic units \cite{Weinhold12}. Equation \ref{20} uses
the notation of Ilkov, et al., \cite{Ilkov92}, but corrects algebraic errors
in its Eq.~8. The functions $f\left( l,m\right) $, \ $A_{m}\left( \omega
,\gamma \right) $,\ and $g\left( \gamma \right) $ are defined therein.

\begin{figure}[H]\centering\includegraphics[width=3.375in]{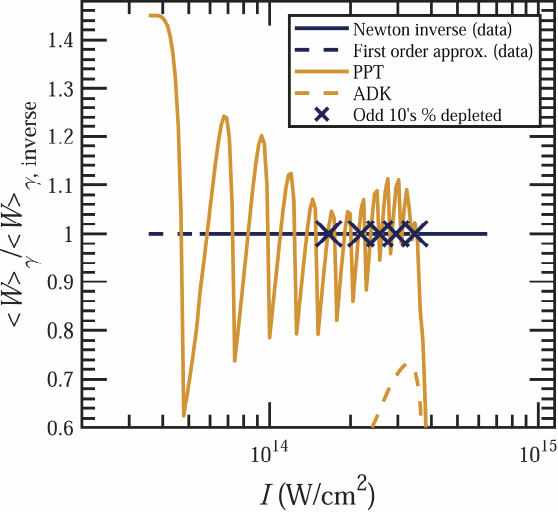}\end{figure}  
\textbf{Fig.~5} The ratio of the PPT and ADK models to the best fitting
inverted Ar$^{+}$ solution to $\left\langle W\right\rangle _{\gamma }$ vs.\ $%
I $ are plotted. The inversion fails before the ADK\ model reaches the
inversion, causing the ratio to collapse. Again, the crosses are at odd 10's
percent peak neutral depletion.

\bigskip

Ilkov derives an analytic approximation to coefficient $\left\vert
C_{n^{\ast }l^{\ast }}^{\ }\right\vert ^{2}$, correcting an error for this
term in Ammosov \cite{Ammosov86}. We generalize the factorial terms of
Ilkov's Eq.~3 for noninteger arguments (as done elsewhere \cite{Song09}) by
expressing them in terms of the gamma function, instead of approximating
them via the Sterling formula as Ilkov does. The result is $\left\vert
C_{n^{\ast }l^{\ast }}^{\ }\right\vert ^{2}=2^{2n^{\ast }}/\left( n^{\ast
}\Gamma \left( 2n^{\ast }\right) \right) $. $U_{0}=15.759$ eV \cite%
{Minnhagen73} and $C_{n^{\ast }l^{\ast }}^{\ }=\allowbreak 2.0289$ for Ar $%
\rightarrow $ Ar$^{+}+$ e$^{-}$, and $\hbar \omega _{0}=1.55$ eV for Guo's $%
800$ nm laser. Being a noble gas, we have atomic orbital angular momentum
and projection quantum numbers $l=0$ and $m=0$, respectively. Figure~5 plots
the ratio of the PPT model for Ar$^{+}$ to the inverted solution to the
data, where $\beta $ and $\kappa n_{0}$ have been adjusted for the best fit
to PPT short of 90\% neutral depletion at peak focus. The ADK model, which
approaches Eq.~\ref{20} in the limit of $\gamma \rightarrow 0$, is overlaid
for reference in Fig.~6, where we use the above $\left\vert C_{n^{\ast
}l^{\ast }}^{\ }\right\vert ^{2}$ in Ammosov's Eq.~1. For this,%
\begin{equation}
\begin{tabular}{l}
$w\left( F\right) =\left( \frac{6}{\pi }\right) ^{1/2}\left\vert C_{n^{\ast
}l^{\ast }}^{\ }\right\vert ^{2}f\left( l,m\right) E_{0}$ \\ 
$\times $ $\left( \frac{2F_{0}}{F}\right) ^{2n^{\ast }-\left\vert
m\right\vert -3/2}\exp \left( -\frac{2F_{0}}{3F}\right) $%
\end{tabular}%
\ \ \ \   \label{21}
\end{equation}

\section{Results}

The best fit of Ar$^{+}$ to PPT occurs for $\alpha =3.762$ and $\beta =0.98$%
. Guo's maximum value of $I_{0}$ (rescaled by $\beta $) is then, $I_{0,\max
}=7.78\times 10^{14}$ W/cm$^{2}$. Eq.~\ref{6} implies the beam half-width is 
$w_{0}=32.0$ $\mu $m and the Rayleigh range is $z_{\mathrm{R}}=4.0$ mm. $%
\kappa n_{0}$ is $2.347,$ $2.939,$ and $3.157\times 10^{20}$ (shot-Torr-m$%
^{3}$)$^{-1}$ for Ar$^{+}$, O$_{2}^{+}$, and N$_{2}^{+}$, respectively.

This $\alpha $ implies, consistent with Eq.~\ref{1}, 
\begin{equation}
\sigma _{8}=\alpha \sigma _{8\mathrm{,Sharma}}=\allowbreak 1.24\times
10^{-129}\mathrm{{\ W}^{-8}{m}^{16}{s}^{-1}}  \label{23}
\end{equation}%
It also implies that the momentum transfer collision frequency for Sharma's
experiment is $\nu _{\mathrm{c}}=$ $\alpha \nu _{\mathrm{c,Sharma}%
}=1.949\times 10^{12}$ s$^{-1}$ \cite{Sharma18}. Given background neutral
air density $n_{\mathrm{air}}=2.44\times 10^{25}$ m$^{-3}$, the implied
temperature is $T=0.8$ eV \cite{Pusateri15} for electrons released by O$_{2}$
exposed to an optical pulse with Sharma's $I_{0}=26.8$ TW/cm$^{2}$ \cite%
{Sharma18}. This $\sigma _{8}$ estimate, being only 23\% higher than the
lower bound established in Sec.~II, suggests that whatever peak is rising at
the end of Kloda's spectrum \cite{Kloda10} is relatively minor and/or the
electrons rescattered off their parent ion to directions \emph{not} recorded
by Kloda do not substantially increase the mean electron kinetic energy.

Figure~6 plots $\left\langle W\right\rangle _{\gamma }$ results of the
inversion algorithm over the range of $I_{0}$ for which\ Guo presents data
(extrapolated ranges excluded). $\kappa N_{\mathrm{R}}\left( I_{0}\right)
/\left( \kappa N_{\mathrm{fit}}\left( I_{0}\right) \right) $, as illustrated
in Fig.~3 for Ar$^{+}$, deviates by less than 0.1\% from unity for the
entire range plotted for all species (results for which the algorithm fails
are not shown). Nonetheless, $\left\langle W\right\rangle _{\gamma }$ for
inverted Ar$^{+}$ and O$_{2}^{+}$ data exhibit an unphysical upward surge
when $I$ exceeds a value where $n/n_{0}\gtrsim 0.9$. This is a result of the
algorithm becoming unstable beyond 90\% depletion, despite producing a
verified solution based on the 0.1\% standard.

\begin{figure}[H]\centering\includegraphics[width=3.375in]{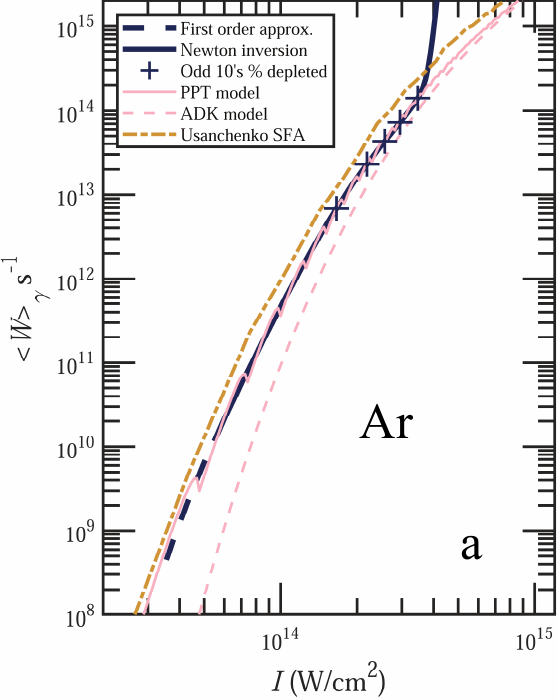}\end{figure} 
\begin{figure}[H]\centering\includegraphics[width=3.375in]{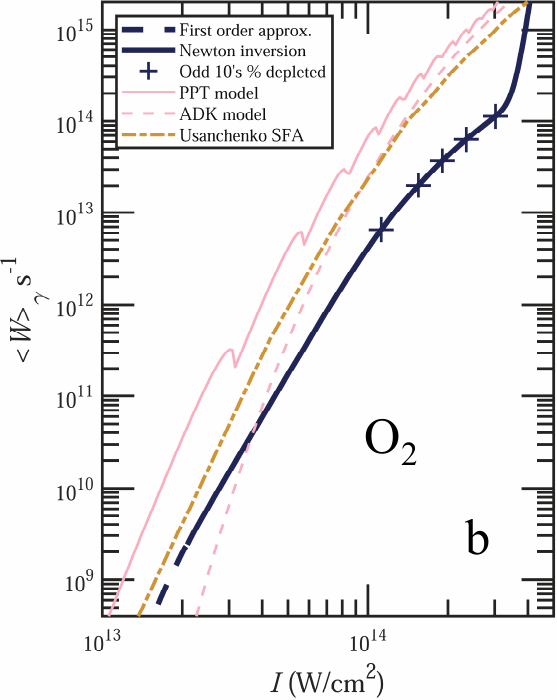}\end{figure} 
\begin{figure}[H]\centering\includegraphics[width=3.375in]{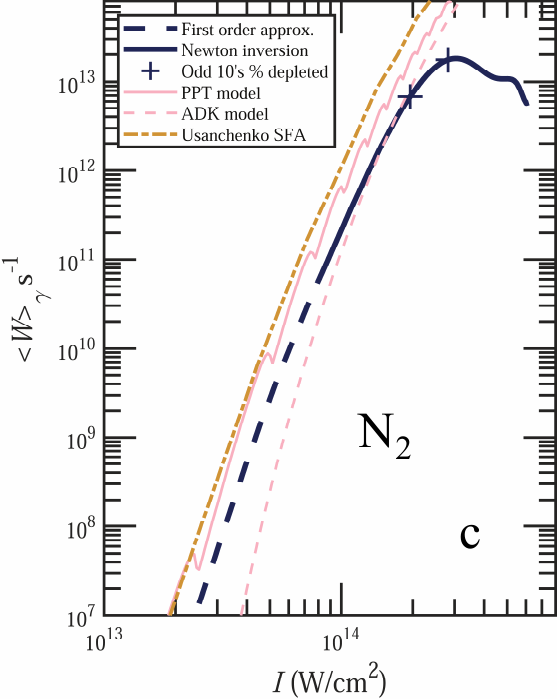}\end{figure} 
\textbf{Fig.~6} Optical-cycle averaged ionization rates based on functional
Newton's method $\left\langle W\right\rangle _{\gamma ,\mathrm{Newt}}$ \ for
Ar, O$_{2}$, and N$_{2}$, with crosses at odd 10's percent neutral
depletion. The method generally fails beyond 90\% depletion. It results in
an unphysical drop in $\left\langle W\right\rangle _{\gamma ,\mathrm{Newt}}$
for N$_{2}$ beyond 20\% depletion, though, likely due to detector saturation
reported for those measurements. The results of the PPT and ADK models, and
the strong field ionization model of Usachenko and Chu, 2005 \cite%
{Usachenko05} are shown for reference.

\bigskip

Reducing $\beta $ to $0.97$, and increasing $\alpha $ to $3.901$ to obtain
the best inverted Ar$^{+}$ data fit to the PPT model for that $\beta $, (but
not as good as for $\beta =0.98$) results in an unphysical percipitous \emph{%
drop} in $\left\langle W\right\rangle _{\gamma }$ for Ar beyond 90\%
depletion. That is, small changes in $\beta $ and refitting\ $\alpha $
results in large changes in the inverted solution beyond this point. $%
\left\langle W\right\rangle _{\gamma }$ for N$_{2}$, however, dips
unphysically beyond only 20\% depletion in Fig.~6c. This is attributed to
detector saturation reducing the MCP signal, as reported by Guo.

Usachenko and Chu, 2005 \cite{Usachenko05} specifically references Guo as
presenting experimental results to which its strong-field ionization model
should be compared, so its results are overlaid in Fig.~6.

\section{Discussion/Conclusions}

It is found that inversion parameters $\alpha $, $\beta $, $N_{\mathrm{box}}$%
, the number of Newton iterations, and the ranges that $I_{0}$ must be
extrapolated\ above and below the data must be adjusted significantly by
trial and error to obtain a solution to the ratio $\kappa N_{\mathrm{R}%
}\left( I_{0}\right) /\left( \kappa N_{\mathrm{fit}}\left( I_{0}\right)
\right) $ that is within 0.1\% of unity, as plotted in Fig.~3 for Ar$^{+}$.
There is an optimum number of iterations beyond which this ratio diverges
from unity for Ar$^{+}$ and O$_{2}^{+}$ inversion ($21$ and $22$,
respectively), for example. The ratio for N$_{2}^{+}$, however, continues to
improve without bound, albiet slowly ($201$ iterations are used).

The assumption that Ar $\rightarrow $\ Ar$^{+}+$ e$^{-}$ is well-represent
by PPT is a possible source of systematic error. A measurement of the
absolute collection efficiency of each of the species of interest, and a
direct measurement of the peak USPL\ pulse intensity would eliminate the
need for this assumption. It is tempting to adjust its invertion's fit away
from the best fit to PPT such that $\beta $ is transitional between causing
an upward and a downward surge in $\left\langle W\right\rangle _{\gamma ,%
\mathrm{Newt}}$ following 90\% neutral depletion. However, the transition
also depends on $\alpha $, and the solution, being strongly dependent on the
details of the laser's far off-peak space-time profile, may not be precisely
Gaussian. So, it is unclear if this will increase accuracy. From Chin 2004 
\cite{Chin04} , \textquotedblleft In this saturation region, no interaction
physics can be extracted ...\textquotedblright .

The MCP alloy differing from NiCr of Meier is another possible source of
error. Meier's Table 3, though, suggests inaccuracy due to use of a
different electrode material is probably not significant (even if it were a
Cu alloy), given that all values of $\mu _{\mathrm{r}}$ for the species of
interest are similar. Nonetheless, suggested improvements leading to more
accurate $\left\langle W\right\rangle _{\gamma }$ results include repeating
the experiments with\ a TOF\ detector\ MCP that employs an alloy for which
collection efficiency data is confirmed to use the same materials, or even
the same device. Ensuring\ the\ spatial-temporal $I$ profile is
well-represented by a Gaussian distribution or generalizing the inversion
algorithm to account for the actual profile, and generalizing the inversion
to account for other ion species would also improve accuracy. Investigating
alternatives to the simple boxcar averaging used by the algorithm to improve
its convergence behavior is also suggested.

\bigskip

\textbf{Funding. \ }This material is based on work supported by Air Force
Office of Scientific Research award FA9550-19RDCOR027.

\textbf{Acknowledgment. \ }The author thanks Jennifer A. Elle and Travis
Garrett for useful conversations.

\textbf{Disclosures. \ }The authors declare no conflicts of interest.

\textbf{Disclaimer. \ }The views expressed are those of the author and do
not necessarily reflect the official policy or position of the Department of
the Air Force, the Department of Defense, or the U.S. government.

\textbf{Data availability. \ }Data underlying the results presented in this
paper are available in the cited references.

\textbf{Release approval. \ }Approved for public release; distribution is
unlimited. Public Affairs release approval \#AFRL-2024-4962.

\bigskip

\newif\ifabfull\abfulltrue

\end{document}